\documentclass[12pt]{article}
\usepackage{graphicx}
\usepackage{lscape}
\usepackage{amssymb,amsmath,slashed}
\usepackage{appendix}
\usepackage{multirow}
\usepackage{amssymb,epsfig}
\usepackage{psfrag}
\usepackage{enumerate,float}
\usepackage{hyperref}

\newcommand{\pp}[1]{\left(#1\right)}
\newcommand{\bb}[1]{\left[#1\right]}
\newcommand{\cc}[1]{\left\{#1\right\}}

\allowdisplaybreaks
\begin{document}
\def\qq{\langle \bar q q \rangle}
\def\uu{\langle \bar u u \rangle}
\def\dd{\langle \bar d d \rangle}
\def\sp{\langle \bar s s \rangle}
\def\GG{\langle g_s^2 G^2 \rangle}
\def\Tr{\mbox{Tr}}
\def\figt#1#2#3{
        \begin{figure}
        $\left. \right.$
        \vspace*{-2cm}
        \begin{center}
        \includegraphics[width=10cm]{#1}
        \end{center}
        \vspace*{-0.2cm}
        \caption{#3}
        \label{#2}
        \end{figure}
    }

\def\figb#1#2#3{
        \begin{figure}
        $\left. \right.$
        \vspace*{-1cm}
        \begin{center}
        \includegraphics[width=10cm]{#1}
        \end{center}
        \vspace*{-0.2cm}
        \caption{#3}
        \label{#2}
        \end{figure}
                }

\newcommand{\V}{{\cal V}}
\newcommand{\A}{{\cal A}}
\newcommand{\T}{{\cal T}}
\def\ds{\displaystyle}
\def\beq{\begin{equation}}
\def\eeq{\end{equation}}
\def\bea{\begin{eqnarray}}
\def\eea{\end{eqnarray}}
\def\beeq{\begin{eqnarray}}
\def\eeeq{\end{eqnarray}}
\def\ve{\vert}
\def\vel{\left|}
\def\ver{\right|}
\def\nnb{\nonumber}
\def\ga{\left(}
\def\dr{\right)}
\def\aga{\left\{}
\def\adr{\right\}}
\def\lla{\left<}
\def\rra{\right>}
\def\rar{\rightarrow}
\def\lrar{\leftrightarrow}
\def\nnb{\nonumber}
\def\la{\langle}
\def\ra{\rangle}
\def\ba{\begin{array}}
\def\ea{\end{array}}
\def\tr{\mbox{Tr}}
\def\ssp{{\Sigma^{*+}}}
\def\sso{{\Sigma^{*0}}}
\def\ssm{{\Sigma^{*-}}}
\def\xis0{{\Xi^{*0}}}
\def\xism{{\Xi^{*-}}}
\def\qs{\la \bar s s \ra}
\def\qu{\la \bar u u \ra}
\def\qd{\la \bar d d \ra}
\def\qq{\la \bar q q \ra}
\def\gGgG{\la g^2 G^2 \ra}
\def\q{\gamma_5 \not\!q}
\def\x{\gamma_5 \not\!x}
\def\g5{\gamma_5}
\def\sb{S_Q^{cf}}
\def\sd{S_d^{be}}
\def\su{S_u^{ad}}
\def\sbp{{S}_Q^{'cf}}
\def\sdp{{S}_d^{'be}}
\def\sup{{S}_u^{'ad}}
\def\ssp{{S}_s^{'??}}

\def\sig{\sigma_{\mu \nu} \gamma_5 p^\mu q^\nu}
\def\fo{f_0(\frac{s_0}{M^2})}
\def\ffi{f_1(\frac{s_0}{M^2})}
\def\fii{f_2(\frac{s_0}{M^2})}
\def\O{{\cal O}}
\def\sl{{\Sigma^0 \Lambda}}
\def\es{\!\!\! &=& \!\!\!}
\def\ap{\!\!\! &\approx& \!\!\!}
\def\md{\!\!\!\! &\mid& \!\!\!\!}
\def\ar{&+& \!\!\!}
\def\ek{&-& \!\!\!}
\def\kek{\!\!\!\!&-& \!\!\!\!}
\def\cp{&\times& \!\!\!}
\def\se{\!\!\! &\simeq& \!\!\!}
\def\eqv{&\equiv& \!\!\!}
\def\kpm{&\pm& \!\!\!}
\def\kmp{&\mp& \!\!\!}
\def\mcdot{\!\cdot\!}
\def\erar{&\rightarrow&}


\def\simlt{\stackrel{<}{{}_\sim}}
\def\simgt{\stackrel{>}{{}_\sim}}


\title{
         {\Large
                 {\bf
                     $ N^*(1535) \to N $ transition form-factors due to the axial current
                 }
         }
      }

\author{\vspace{1cm}\\
{\small T. M. Aliev\,$^a\!\!$ \thanks {e-mail:
taliev@metu.edu.tr}\,\,,
T. Barakat\,$ ^b\!\! $ \thanks{email: tbarakat@ksu.edu.sa}\,\,,
K. \c Sim\c sek\,$^a\!\!$ \thanks {e-mail:
ksimsek@metu.edu.tr}\,\,, 
} \\
{\small $^a$ Physics Department, Middle East Technical University,
06531 Ankara, Turkey} \\
{\small $ ^b $ Physics Department, King Saud University, Riyadh 11451, Saudi Arabia}
}
\date{}

\begin{titlepage}
\maketitle
\thispagestyle{empty}

\begin{abstract}
\baselineskip 0.5 cm
The form-factors for the transition $ N^*(1535)\to N $ induced by isovector and isoscalar axial currents within the framework of light-cone QCD sum rules by using the most general form of the interpolating current are calculated. In numerical calculations, we use two sets of values of input parameters. It is observed that the $ Q^2 $ dependence of the form-factor $ G _A $ can be described by the dipole form. Moreover, the form-factors $ G _P^{(S)} $ are found to be highly sensitive to the variations in the auxiliary parameter $ \beta $.
\end{abstract}

\end{titlepage}

\section{Introduction}
Nucleon form-factors are fundamental quantities for understanding the inner structure of hadrons at low energies. The electromagnetic form-factors of nucleon are studied in a wide range of momentum transfer square (see \cite{r1}). The electromagnetic form-factors within the light-cone sum rules (LCSR) were comprehensively studied in many works (see for example \cite{r2,r3,r4,r5,r6}). Unlike the electromagnetic form-factors, those induced by isovector and isoscalar axial currents are not measured. Only the nucleon axial charge is experimentally well determined from the neutron $ \beta $ decay, and the latest value is $ g _A = 1.2724 $ \cite{r7}. The determination of the axial form-factors can give very useful information on the flavor structure and spin content of nucleon resonances. Therefore, the study of these form-factors receives special attention for understanding the structure of nucleon resonances.\par
	 The LCSR results for the nucleon axial form-factors and for tensor factors are studied in \cite{r8,r9} and \cite{r10,r11}, respectively. The axial form-factors of the nucleon within the holography approach are investigated in \cite{r12}. In the lattice QCD, the momentum-transfer dependence of the axial form-factor is studied in \cite{r13,r14,r15,r16,r17}. The study of the momentum transfer dependence of the axial form-factor is possible only in pion electro-production and neutrino-induced charged-current reactions.\par 
	The experiments planned at Jefferson Laboratory (JLab), MAMI, are aimed to study properties of baryon resonances in photo- and electro-production reactions \cite{r18}.\par 
	Motivated by the prospective experiments at JLab, MAMI, we aim to study the form-factors for the transition $ N^*(1535) \to N $ induced by isovector and isoscalar axial currents in the framework of the LCSR. Note that the electromagnetic form-factors for the transition $ N^*(1535)\to N $ within the LCSR are calculated in \cite{r19}.\par
	 
	This paper is structured as follows. In Section 2, we introduce the relevant correlation function and derive the LCSR for the form-factors induced by axial quark current. Our numerical analysis for axial form-factors for the transition $ N^*(1535) \to N $ is presented in Section 3. This section also contains our discussion and conclusion.  

\section{Sum rules for the transition $ N^*(1535)\to N $ form-factors induced by axial current}
	In this section, we derive the LSCR for the form-factors for the transition $ N^*(1535) \to N $ induced by isovector and isoscalar axial currents. To this end, we introduce the following vacuum to $ N^*(1535) $ correlation function:
	\begin{align}
		\Pi _\mu (p,q) = i \int d^4x \ e^{iq\cdot x} \langle 0 \vert T \{ \eta (0) A _\mu ^{(S)} (x) \} \vert N^* (p) \rangle. \label{e1}
	\end{align}
	In Eq. \eqref{e1}, $ \eta $ is the interpolating current for the nucleon and $ A _\mu^{(S)} $ is the isovector (isoscalar) axial vector current,
	\begin{align}
		A _\mu^{(S)} = \bar u \gamma _\mu \gamma _5 u \mp \bar d \gamma _\mu \gamma _5 d. \label{e2}
	\end{align}
	The nucleon interpolating current, in general, can be written as
	\begin{align}
		\eta = 2 \varepsilon ^{abc} \sum _{\ell =1}^2 (u^{aT} C A _1^\ell d ^b) A _2^\ell u^c \label{e3}
	\end{align}
	where $ a $, $ b $, and $ c $ are color indices, $ C $ is the charge-conjugation operator, $ A _1^1 = I $, $ A _2^1 = \gamma _5 $, $ A _1^2 = \gamma _5 $, and $ A _2^2 = \beta I $. The case $ \beta = -1 $ corresponds to the Ioffe current.\par 
	The LCSR for the relevant quantities is obtained by calculating the correlation function from the hadronic and QCD sides and then matching the result of two representations. From the hadronic side, the result can be obtained by inserting the total set of the nucleon states between the nucleon interpolating current and the axial current $ A _\mu^{(S)} $, and separating the contribution of the ground-state nucleon. \par 
	The standard procedure for obtaining QCD sum rules is the calculation of the correlation function from hadronic and QCD sides. We begin with the calculation of the hadronic side of the correlation function \eqref{e1}. \par
	Now we will give a few details of this procedure. We explicitly keep the contribution of the state $ N(940) $ in the hadronic dispersion relation and, according to the quark-hadron duality ans\"atz, represent higher states by a dispersion integral, starting from some threshold, $ s _0 $. \par 
	After separating the contribution from the ground-state $ N $ in the hadronic part, we get
	\begin{align}
		\Pi _\mu = \frac{\langle 0 \vert \eta \vert N \rangle \langle N \vert A _\mu^{(S)} \vert N^* \rangle}{m^2 - p'^2} + \cdots \label{e4}
	\end{align}
	where $ \cdots $ denotes the contributions from higher states and the continuum. The coupling of $ N $ to the interpolating current $ \eta $, that is, the decay constant or the residue, is defined as
	\begin{align}
		\langle 0 \vert \eta \vert N (p) \rangle = \lambda _Nu _N (p). \label{e5}
	\end{align}
	The matrix element of the axial current between $ N $ and $ N^* $ in terms of three form-factors is given by
	\begin{align}
		\langle N(p') \vert A _\mu \vert N^*(p) \rangle &= \bar u _N (p') \Big[
			\gamma _\mu \gamma _5 G _A(q^2) + \frac{q _\mu}{m _1+m _2} \gamma _5 G _P(q^2) \nonumber\\&+i \sigma _{\mu\nu} \frac{q _\nu}{m _1+m _2} \gamma _5 G _T (q^2) 
		\Big]\gamma _5 u _{N^*} (p) \label{e6}
	\end{align}
	where $ G _A $, $ G _P $, and $ G _T $ are the axial, induced pseudoscalar, and induced tensor form-factors, respectively. The matrix element of the isoscalar axial-vector current between $ N $ and $ N^* $ is obtained from \eqref{e6} by replacing $ A _\mu \to A _\mu^S $, $ G _A \to G _A^S $, $ G _P\to G _P^S $, and $ G _T\to G _T^S $. The $ G $-parity invariance and taking the exact isospin symmetry limit lead to the result that $ G _T $ and $ G _T^S $ form-factors must vanish. Therefore, in all the following discussions these form-factors are taken to be zero. Taking into account this fact and putting Eqs. \eqref{e6} and \eqref{e5} into \eqref{e4} for the hadronic contributions to the correlation function, we obtain
	\begin{align}
		\Pi _\mu = \frac{\lambda _N}{m _N^2 - p'^2} (\slashed p-\slashed q + m _N) \Big[
			\gamma _\mu \gamma _5 G _A + \frac{q _\mu}{m _N+ m _{N^*}} \gamma _5 G _P(q^2) 
		\Big]\gamma _5 u _{N^*} (p) + \cdots \label{e7}
	\end{align}
	Using the equation of motion, namely $ \slashed p u _{N^*} (p) = m _{N^*} u _{N^*} (p) $, in the correlation function from hadronic part, we get
	\begin{align}
		\Pi _\mu &= \frac{\lambda _N}{m _N^2 - p'^2 } \Big\{  
			(m _N - m _{N^*}) \gamma _\mu  G _A + 2 p _\mu  G _A -  \slashed q \gamma _\mu  G _A \nonumber\\&+ \frac{ G _P q _\mu}{m _{N^*} + m _N} (m _{N^*} + m _N - \slashed q)
		\Big\} + \cdots \label{e8}
	\end{align}
	where $ p'=p-q $. From this expression, we see that the correlator function can be decomposed into the following Lorentz structures:
	\begin{align}
		\Pi _\mu (p,q) &= \Pi _1 p _\mu + \Pi _2 \gamma _\mu +\Pi _3 \slashed q \gamma _\mu +\Pi _4 q _\mu +\Pi _5 q _\mu \slashed q. \label{e9}
	\end{align}
	Now, let us turn our attention to the calculation of the correlation function from the QCD side. For $ p'^2,q^2 \ll 0 $, the product of the two currents in Eq. \eqref{e1} can be expanded around the light cone, $ x^2 \sim 0 $. The result of the operator-product expansion (OPE) is obtained by a sum over the distribution amplitudes (DAs) of $ N^* $, with an increasing twist multiplied by their corresponding coefficient functions.\par 
	At this point, we would like to present some details of the calculations. Using the forms of interpolating current and axial current given above, we get
	\begin{align}
		(\Pi _\mu) _\rho &= \frac{1}{2} \int d^4 x \ e^{iq\cdot x} \sum _{\ell =1}^2 \Big\{
			(C A _1^\ell) _{\alpha\gamma} [A _2^\ell S _u(-x) \gamma _\mu \gamma _5] _{\rho\beta} \nonumber\\ & + (A _2^\ell) _{\rho\alpha} [(C A _1^\ell) ^T S _u(-x) \gamma _\mu \gamma _5]_{\gamma\beta} \nonumber\\ & \mp (A _2^\ell) _{\rho \beta} [C A _1^\ell S _u(-x) \gamma _\mu \gamma _5] _{\alpha\gamma}  
		\Big\} 4 \varepsilon ^{abc} \langle 0 \vert u _\alpha^a (0) u _\beta^b (x) d _\gamma^c (0) \vert N^* (p) \rangle. \label{e10}
	\end{align}
	In Eq. \eqref{e10}, the upper (lower) sign in the last line corresponds to the isovector (isoscalar) axial current, and $ S _q(x) $ is the light-quark propagator. The light-quark propagator in the presence of external electromagnetic and gluonic background fields were calculated in \cite{r20} and are given by
	\begin{align}
		S _q(x) &= \frac{i\slashed x }{2\pi^2 x^4} [x,0] - \frac{ig}{16\pi^2x^2} \int _0^1 du [u,ux] \{ \bar u \slashed x \sigma _{\alpha\beta} + u \sigma _{\alpha\beta} \slashed x \} G ^{\alpha\beta} (ux) [ux,0] \nonumber \\ & - \frac{ig}{16\pi^2x^2} \int _0^1 du [u,ux] \{ \bar u \slashed x \sigma _{\alpha\beta} +u\sigma _{\alpha\beta} \slashed x \} F^{\alpha\beta} (ux) [ux,0]. \label{e11}
	\end{align}
	In Eq. \eqref{e11},
	\begin{align*}
		[x,y] := P \exp \Big\{
			i \int _0^1 dt (x-y) _\mu [e _q A ^\mu (tx-\bar t y) + g B^\mu (tx-\bar ty)]
		\Big\}
	\end{align*}
	is the path-ordering exponent. In the calculations, we use the Fock-Schwinger gauge, i.e. $ x _\mu A^\mu (B^\mu) = 0 $, hence these factors can be safely omitted.\par 
	The matrix element of the three quarks between the vacuum and the states $ N^* $, defined in terms of the DAs of $ N^* $ with increasing twist, is given in \cite{r21}. For the sake of completeness, we present these DAs in Appendix A.\par 
	Using the explicit expressions of the DAs of $ N^* $, performing the integration over $ x $, and selecting the coefficients of the structures $ \slashed q\gamma _\mu $ and $ \slashed q q _\mu $ in both representations, we get
	\begin{align}
		\begin{split}
		- \frac{\lambda _N}{m _N^2 - p'^2}  G _A ^{(S)} & = \Pi _3^{(S)},\\
		- \frac{\lambda _N}{m _N^2 - p'^2} \frac{G _P^{(S)}}{m _{N^*} + m _N} & = \Pi _5 ^{(S)}.
		\end{split}\label{e12}
	\end{align} 
	In Eq. \eqref{e12}, $ \Pi _5 ^{(S)} $ and $ \Pi _3 ^{(S)} $ are given by
\def\mnstar{m _{N^*}}
\newcommand*{\ttilde}[1]{\tilde{\tilde #1}}
	\begin{align}
	\Pi _5 ^{(S)} &= \mnstar ^2 \int _0^1 dx _2 \cc{\frac{(1-\beta) N _1(x _2) + (1+\beta) N _2 (x _2)}{\Delta ^2(x _2)}+ \frac{(1+\beta)N _3(x _2)}{\Delta ^3 (x _2)}}\nonumber\\
	& \mp \mnstar ^2 \int _0^1 dx _3 \cc{
		\frac{(1-\beta) N _4 (x _3) + (1+\beta) N _5(x _3)}{\Delta ^2 (x _3)} + \frac{(1+\beta)N _6(x _3)}{\Delta ^3 (x _3)}
	}\label{e13}
	\end{align}
	and
	\begin{align}
	\Pi _3 ^{(S)} &= \mnstar 
	\int _0^1 dx _2 \cc{
		\frac{(1-\beta)N _7 (x _2) + (1+\beta) N _8(x _2)}{\Delta (x _2)} + \frac{(1-\beta) N _9(x _2) + (1+\beta) N _{10} (x _2)}{\Delta ^2 (x _2)}
	}\nonumber\\
	& \mp \mnstar \int _0^1 dx _3 \cc{
		\frac{(1-\beta) N _{11}(x _3)+ (1+\beta) N _{12} (x _3)}{\Delta (x _3)} + \frac{(1-\beta) N _{13}(x _3)+ (1+\beta) N _{14} (x _3)}{\Delta ^2 (x _3)}
	}\label{e14}
	\end{align}
where 
\begingroup\allowdisplaybreaks
\begin{align}
	N _1 (x) & = - \tilde A _{123} (x) + \tilde A _{1345} (x) - 2\tilde A _{34} (x) - \tilde V _{123} (x) + \tilde V _{34} (x), \nonumber\\
	N _2 (x) &= - \tilde P _ {12} (x) + \tilde S _{12} (x) - 2 \tilde T _{123} (x) - 3 \tilde T _{127} (x) - 5 \tilde T _{158} (x) - 10 \tilde T _{78} (x) +\frac{2}{x} \ttilde T _{234578} (x),\nonumber \\
	N _3 (x) &= \frac{2}{x}\pp{Q^2 + \mnstar ^2x^2} \ttilde T _{234578} (x),\nonumber \\
	N _4 (x) &= \tilde A _{123} (x) + \tilde A _{34} (x) + \tilde V _{123} (x) - \tilde V _{34} (x),\nonumber\\
	N _5 (x) &= -\tilde P _{12} (x) + \tilde S _{12} (x) - \tilde T _{127} (x) - \tilde T _{158} (x) - 2 \tilde T _{78} (x) + \frac{2}{x} \ttilde T _{234578} (x),\nonumber \\
	N _6 (x) &= \frac{2}{x} \pp{Q^2 + \mnstar^2 x^2} \ttilde T _{234578} (x),\nonumber \\
	N _7(x) &= - A _3 (x) + V _1 (x) - V _3 (x) - \frac{1}{2x} \bb{\tilde A _{123} (x) +  \tilde V _{123}(x)}, \label{e15}\\
	N _8(x) &= P _1 (x) + S _1 (x) + 2 T _1 (x) - 4 T _7 (x) -  \frac{1}{2x} \bb{\tilde T _{123} (x) + 3 \tilde T _{127} (x)}  , \nonumber \\
	N _9 (x) &= -\frac{1}{2x} \pp{Q^2 + \mnstar ^2 x^2} \bb{\tilde A _{123}(x) + \tilde V _{123} (x)}  - \mnstar ^2 \bb{\ttilde A _{123456} (x) - \ttilde V _{123456} (x)},\nonumber  \\
	N _{10} (x) &= -\frac{1}{2x} \pp{Q^2 + \mnstar ^2 x^2}\bb{\tilde T _{123} (x) + 3  \tilde T _{127} (x)}  + \mnstar^2 \ttilde T _{234578} (x),\nonumber\\
	N _{11} (x) &= \frac{1}{2}\bb{A _1 (x) -  V _1 (x)},\nonumber \\
	N _{12} (x) &= \frac{1}{2}\bb{ P _1 (x) +  S _1 (x) - T _1 (x) + 2  T _7 (x)}    +\frac{1}{4x} \bb{\tilde T _{123} (x) + \tilde T _{127} (x)} ,  \nonumber\\
	N _{13} (x) &= -\mnstar ^2\bb{\ttilde A _{123456} (x) + \ttilde V _{123456} (x)},  \nonumber\\
	N _{14} (x) &=  \frac{1}{4x} \pp{Q^2 + \mnstar ^2 x ^2} \bb{\tilde T _{123} (x) +  \tilde T _{127} (x)}   + \frac{\mnstar^2}{4} \ttilde T _{234578} (x),\nonumber
\end{align}
\endgroup 
and
\begingroup\allowdisplaybreaks
\begin{align}
	\tilde S _{12} &=\tilde S _1 -\tilde S _2,\nonumber\\
	\tilde P _{12} &=\tilde P _2 -\tilde P _1,\nonumber\\
	\tilde V _{123} &=\tilde V _1 -\tilde V _2 -\tilde V _3,\nonumber\\
	\tilde V _{1345} &= -2\tilde V _1 +\tilde V _3 +\tilde V _4 + 2\tilde V _5,\nonumber\\
	\tilde V _{34} &=\tilde V _4 -\tilde V _3,\nonumber\\
	\tilde{\tilde V} _{123456} &= -\ttilde V _1 +\ttilde V _2 +\ttilde V _3 +\ttilde V _4 +\ttilde V _5 -\ttilde V _6,\nonumber\\
	\tilde A _{123} &= -\tilde A _1 +\tilde A _2 -\tilde A _3,\nonumber\\
	\tilde A _{1345} &= -2\tilde A _1 -\tilde A _3 -\tilde A _4 + 2\tilde A _5 ,\label{e16}\\
	\tilde A _{34} &=\tilde A _3 -\tilde A _4, \nonumber\\
	\tilde{\tilde A} _ {123456} &=\ttilde A _1 -\ttilde A _2 +\ttilde A _3 +\ttilde A _4 -\ttilde A _5 +\ttilde A _6,\nonumber\\
	\tilde T _{123} &=\tilde T _1 +\tilde T _2 - 2\tilde T _3,\nonumber\\
	\tilde T _{127} &=\tilde T _1 -\tilde T _2 - 2\tilde T _7,\nonumber\\
	\tilde T _{158} &= -\tilde T _1 +\tilde T _5 + 2\tilde T _8,\nonumber\\
	\tilde{\tilde T}  _{234578} &= 2\ttilde T _2 - 2\ttilde T _3 - 2\ttilde T _4 + 2\ttilde T _5 + 2\ttilde T _7 + 2\ttilde T _8,\nonumber\\
	\tilde T _{78} &=\tilde T _7 -\tilde T _8,\nonumber\\
	\tilde{\tilde T} _{125678} &= -\ttilde T _1 +\ttilde T _2 +\ttilde T _5 -\ttilde T _6 + 2\ttilde T _7 + 2\ttilde T _8.\nonumber
\end{align}
\endgroup
The generic expressions for the DAs with and without tildes are given as 
\begin{align}
	\begin{split}
	F (x _2) &= \int _0 ^{1-x _2} dx _1 \ F (x _1, x _2 , 1- x _1  - x _2), \\
	\tilde F (x _2) &= \int _1^{x _2} dx _2' \int _0^{1-x _2'} dx _1 \ F (x _1, x _2', 1-x _1-x _2'),\\
	\tilde{\tilde F} (x _2) &= \int _1^{x _2} dx _2' \int _1^{x _2'}dx _2''\int _0^{1-x _2''}dx _1\  F (x _1,x _2'',1-x _1-x _2'').
	\end{split}\label{e17}
\end{align}
A similar set of expressions holds true when the argument is $ x _3 $.\par 
Finally, for the derivation of the sum rules for the relevant form-factors, we perform a Borel transformation in the variable $ - (q-p)^2 $ in order to suppress the contributions from higher states and the continuum and to enhance the contribution of the ground state. This can be achieved with the help of the following subtraction rules:
\begin{align}
\begin{split}
\int dx \frac{\rho (x)}{\Delta(x)} & \to -\int _{x _0}^1 \frac{dx}{x} \rho  (x) e ^{-s(x)/M^2},\\
\int dx \frac{\rho  (x)}{\Delta^2(x)} & \to \frac{1}{M^2} \int _{x _0}^1 \frac{dx}{x^2} \rho  (x) e ^{-s(x)/M^2} + \frac{\rho (x _0) e ^{-s _0/M^2}}{x _0^2 m _{N^*}^2 +Q^2}, \\
\int dx \frac{\rho (x)}{\Delta ^3(x)} & \to - \frac{1}{2M^4} \int _{x _0}^1 \frac{dx}{x^3} \rho (x) e^{-s(x)/M^2}- \frac{1}{2x _0} \frac{\rho (x _0)e ^{-s _0/M^2}}{(Q^2+x _0^2 m _{N^*})M^2}\\ & + \frac{1}{2} \frac{x _0^2 e ^{-s _0/M^2}}{Q^2 + x _0^2 m _{N^*}^2} \frac{d}{dx _0} \bb{\frac{1}{x _0} \frac{\rho (x _0)}{Q^2 + x _0^2 m _{N^*}^2}}.
\end{split} \label{e18}
\end{align}
In Eq. \eqref{e18}, the denominator on the left-hand side is defined as
\begin{align*}
\Delta (x):=  (xp -q)^2 = x (p-q)^2 - \bar x Q^2 - x \bar x m _{N^*}^2 
\end{align*}
where $ \bar x = 1-x $. $ x _0 $ is the solution of $ s (x) = s _0 $ where
\begin{align*}
s (x) = \frac{\bar x Q^2 +x \bar x m _{N^*}^2}{x}.
\end{align*}
\section{Numerical analysis}
	Having the explicit expressions of the form-factors $ G _A^{(S)} $ and $ G _P^{(S)} $, now we perform the numerical analysis of the sum rules for them.\par 
	The main non-perturbative input ingredients of the LCSR for the form-factors $ G _A^{(S)} $ and $ G _P^{(S)} $ are the DAs of $ N^* $. They are presented in \cite{r21}. The values of the parameters appearing in the DAs can also be found in \cite{r21}. In numerical calculations, we use two set of values of parameters, i.e. LCSR-1 and LCSR-2. \par 
	The sum rules for the form-factors $ G _A^{(S)} $ and $ G _P^{(S)} $ contain three auxiliary parameters, i.e. the Borel mass squared, $ M^2 $, the continuum threshold, $ s _0 $, and the parameter $ \beta $. The form-factors should be independent of these parameters. Therefore, the primary aim of any sum rules analysis is to find the appropriate domains of these auxiliary parameters for which the form-factors exhibit good stability to their variations. \par 
	The working regions of $ M^2 $ and $ s _0 $ are determined from the standard criterion that the contributions from higher states and the continuum and as well as higher-twist terms should be suppressed. Our analysis shows that the working regions of $ M^2 $ and $ s _0 $, which satisfy the aforementioned criterion are $ 1\ \mathrm{GeV^2} \leq M^2 \leq 3 \ \mathrm{GeV^2} $ and $ s _0 = (2.25 \pm 0.25) \ \mathrm{GeV^2} $. In \cite{r22}, the mass and residues of octet baryons are analyzed, and in the present work, we use the result for the residue of the nucleon given in the same reference. The uncertainties of the input parameters and intrinsic uncertainties carried by the LCSR method (namely the factorization scale, $ M^2 $, higher-twist corrections) can introduce an uncertainty about $ (15\pm 5)\% $. Therefore, we expect that the sum rules for the form-factors work effectively in the domain
	\begin{align}
		2 \ \mathrm{GeV^2} \leq Q^2 \leq 6 \ \mathrm{GeV^2}.\label{e19}
	\end{align}
	Note that the LCSR approach is not dependable for $ Q^2 \leq 1 \ \mathrm{GeV}^2 $. It is because the mass corrections are proportional to $ m _{N}^2/Q^2 $, which becomes very large for $ Q^2 < 1 \ \mathrm{GeV}^2 $, and hence the LCSR becomes unreliable.

\section{Results and discussion}
In Fig.s (1)--(8), we depict the dependence of the form-factors $ G _A^{(S)} $ and $ G _P^{(S)} $ on $ Q^2 $ for various values of $ \beta $ in its working region and at the fixed values of $ M^2 = 2 \ \mathrm{GeV^2} $ and $ s _0 = 2.25 \ \mathrm{GeV^2} $ for two sets of values of input parameters entering the DAs. \par 
From the figures, we make the following observations:
\begin{enumerate}
	[$ \bullet $]
	\item The values of $ G _A $ and $ G _A^S $ are positive (negative) for negative (positive) values of $ \beta $, whereas $ G _P $ and $ G _P^S $ have the same sign as $ \beta $ for both sets of parameters, LCSR-1 and LCSR-2.
	\item The moduli of the form-factors $ G _A $, $ G _A^S $, $ G _P $, and $ G _P^S $ for the second set of values are larger than those for the first set of parameters.
	\item The form-factors $ G _P $ and $ G _P^S $ are sensitive to the variations in $ \beta $ when $ \beta > 0 $ for both sets of input parameters. For example, the values of the form-factors $ G _P $ and $ G _P^S $ at $ \beta = 1 $ are nearly twice as large as the corresponding values of these form-factors for other values of $ \beta > 0 $. 
	\item All the form-factors considered display a similar dependence on $ Q^2 $ for both sets (LCSR-1 and LCSR-2) and for all values of $ \beta $. 
	\item The set LCSR-1 gives the form-factor $ G _P^S $ approximately twice that predicted by the set LCSR-2 for $ \beta < 0 $.
\end{enumerate}
From our results, it follows that the form-factor $ G _A(Q^2) $ can be parametrized by the dipole form
\begin{align}
	G _A (Q^2) = \frac{g _A}{(1+Q^2/m _A^2)^2}\label{e20}
\end{align}
where we tabulated the average of the values of $ g _A $ and $ m _A $ for both parameter sets and for all the values of $ \beta $ that we considered in Table \ref{tab:1}.
\begin{table}
	[H]\centering 
	\caption{The average values of $ g _A $ and $ m _A $ in Eq. \eqref{e20}}\label{tab:1}
	\begin{tabular}
		{cc|c|c|c}
		\hline 
		\hline 
		& \multicolumn{2}{c}{$ g _A $} & \multicolumn{2}{c}{$ m _A \ \mathrm{(GeV)}$}\\
		\hline 
		 & LCSR-1 & LCSR-2 & LCSR-1 & LCSR-2\\
		\hline 
		$ \beta < 0 $ & $ 3.65 \pm 0.70 $&$ 4.55\pm 0.66 $& $ 1.05 \pm 0.02 $ & $ 1.20 \pm 0.02 $\\
		$ \beta > 0 $ & $ -5.82 \pm 0.40 $&$ -6.61 \pm 0.53 $& $ 1.00 \pm 0.02 $& $ 1.16 \pm 0.01 $\\
		\hline 
		\hline 
	\end{tabular}
\end{table}
From Table \ref{tab:1}, it follows that the values of $ g _A $ and $ m _A $ for the LCSR-2 case is slightly larger than the ones for the LCSR-1 one. The magnitude of $ g _A $ for the case $ \beta > 0 $ is larger than that for $ \beta < 0 $. Note that for $ \beta = -1 $ (the case of the Ioffe current), for the values of $ g _A $ and $ m _A $ we find
\begin{align}
	g _A = \begin{cases}
	2.66, & \mbox{LCSR-1},\\
	3.61, & \mbox{LCSR-2},
	\end{cases}
	\quad \mbox{and}\quad 
	m _A = \begin{cases}
	1.08 \ \mathrm{GeV}, & \mbox{LCSR-1},\\
	1.23 \ \mathrm{GeV}, & \mbox{LCSR-2}.
	\end{cases}
\end{align}
We observe that the value of $ m $ is practically insensitive to the variations of both positive and negative values of $ \beta $.\par 
When the experimental results will be obtained, there may appear the possibility to compare our predictions with experimental data.\par 
Finally, we would like to note that our results can be improved by taking into account radiative corrections and the contributions of four and five particles as well as more precise determination of the input parameters.
\section*{Conclusion}
In the present work, we calculated the $ N^*(1535)\to N $ transition form-factors induced by isovector and isoscalar axial currents within the light-cone QCD sum rules using the general form of the interpolating current. In performing numerical analysis, we used two sets of values of input parameters. We obtained that the form-factor $ G _A (Q^2) $ exhibits the dipole form dependence on $ Q^2 $ and we find the values of relevant parameters, namely $ g _A $ and $ m _A $. It was also obtained that $ G _P^{(S)} $ shows a strong dependence on the variations of the auxiliary parameter $ \beta $.
\section{Acknowledgment}
 One of the authors (T. Barakat) extends his appreciation to the Deanship Scientific Research at King Saud University for funding his work through research program RG-1440-090.
  
\clearpage

\section*{Appendix A: $N^\ast$ distribution amplitudes
}

In this Appendix, we present the $N^\ast$ DAs, which are necessary to
calculate the $\Lambda \to N^\ast$ transition form-factors. The DAs of the
$N^\ast$ baryon are defined from the matrix element
$\langle 0 \vel \epsilon^{abc} u_\alpha^a(a_1 x) d_\beta^b(a_2 x)
d_\gamma^c(a_3 x) \ver N^\ast(p)\rangle$. The general decomposition of
this matrix in terms of the DAs of the $N^\ast$ baryon is given below.
(see \cite{r21}),
\bea\label{wave func}
&& 4 \langle 0 \vel \epsilon^{abc} u_\alpha^a(a_1 x) d_\beta^b(a_2 x)
d_\gamma^c(a_3 x) \ver N^\ast(p)\rangle\nnb\\
\es \mathcal{S}_1 m_{N^\ast}C_{\alpha\beta}N_{\gamma}^\ast -
\mathcal{S}_2 m_{N^\ast}^2 C_{\alpha\beta}(\rlap/x N^\ast)_{\gamma}\nnb\\
\ar \mathcal{P}_1 m_{N^\ast}(\gamma_5 C)_{\alpha\beta}(\gamma_5 N^\ast)_{\gamma} +
\mathcal{P}_2 m_{N^\ast}^2 (\gamma_5 C)_{\alpha\beta}(\gamma_5 \rlap/x N^\ast)_{\gamma} -
\left(\mathcal{V}_1 + \frac{x^2 m_{N^\ast}^2}{4} \mathcal{V}_1^M \right) 
(\rlap/p C)_{\alpha\beta} N_{\gamma}^\ast \nnb\\
\ar \mathcal{V}_2 m_{N^\ast}(\rlap/p C)_{\alpha\beta}(\rlap/x N^\ast)_{\gamma} + 
\mathcal{V}_3 m_{N^\ast}(\gamma_\mu C)_{\alpha\beta} (\gamma^\mu N^\ast)_{\gamma} -
\mathcal{V}_4 m_{N^\ast}^2 (\rlap/x C)_{\alpha\beta} N_{\gamma}^\ast \nnb\\
\ek \mathcal{V}_5 m_{N^\ast}^2(\gamma_\mu C)_{\alpha\beta}
(i \sigma^{\mu\nu} x_\nu N^\ast)_\gamma +
\mathcal{V}_6 m_{N^\ast}^3 (\rlap/x C)_{\alpha\beta}(\rlap/x N^\ast)_{\gamma} \nnb \\
\ek \left(\mathcal{A}_1 + \frac{x^2m_{N^\ast}^2}{4}\mathcal{A}_1^M\right)
(\rlap/p \gamma_5 C)_{\alpha\beta} (\gamma N^\ast)_{\gamma} +
\mathcal{A}_2 m_{N^\ast}(\rlap/p \gamma_5 C)_{\alpha\beta} (\rlap/x \gamma_5 N^\ast)_{\gamma} +
\mathcal{A}_3 m_{N^\ast}(\gamma_\mu\gamma_5 C)_{\alpha\beta}
(\gamma^\mu \gamma_5 N^\ast)_{\gamma}\nnb\\
\ek \mathcal{A}_4 m_{N^\ast}^2(\rlap/x \gamma_5 C)_{\alpha\beta}
(\gamma_5 N^\ast)_{\gamma} -
\mathcal{A}_5 m_{N^\ast}^2(\gamma_\mu \gamma_5 C)_{\alpha\beta}
(i \sigma^{\mu\nu} x_\nu \gamma_5 N^\ast)_{\gamma} +
\mathcal{A}_6 m_{N^\ast}^3(\rlap/x \gamma_5 C)_{\alpha\beta}
(\rlap/x \gamma_5 N^\ast)_{\gamma}\nnb\\
\ek \left(\mathcal{T}_1 + \frac{x^2m_{N^\ast}^2}{4}\mathcal{T}_1^M \right)
(i \sigma_{\mu\nu} p_\nu C)_{\alpha\beta} (\gamma^\mu N^\ast)_{\gamma} +
\mathcal{T}_2 m_{N^\ast} (i \sigma_{\mu\nu} x^\mu p^\nu C)_{\alpha\beta}
N_{\gamma}^\ast\nnb\\
\ar \mathcal{T}_3 m_{N^\ast}(\sigma_{\mu\nu} C)_{\alpha\beta}
(\sigma^{\mu\nu} N^\ast)_{\gamma} +
\mathcal{T}_4 m_{N^\ast} (\sigma_{\mu\nu} p^\nu C)_{\alpha\beta}
(\sigma^{\mu\rho} x_\rho N^\ast)_{\gamma} \nnb\\
\ek \mathcal{T}_5 m_{N^\ast}^2 (i\sigma_{\mu\nu} x^\nu C)_{\alpha\beta}
(\gamma^\mu N^\ast)_{\gamma} -  
\mathcal{T}_6 m_{N^\ast}^2 (i \sigma_{\mu\nu} x^\mu p^\nu C)_{\alpha\beta}
(\rlap/x N^\ast)_{\gamma}\nnb\\
\ek \mathcal{T}_7 m_{N^\ast}^2 (\sigma_{\mu\nu} C)_{\alpha\beta}
(\sigma^{\mu\nu} \rlap/x N^\ast)_{\gamma} +
\mathcal{T}_8 m_{N^\ast}^3 (\sigma_{\mu\nu} x^\nu C)_{\alpha\beta}
(\sigma^{\mu\rho} x_\rho N^\ast)_{\gamma}~.\nnb
\eea
The functions labeled with calligraphic letters 
in the above expression do not possess definite twists
but they can be
written in terms of the $N^\ast$ distribution amplitudes (DAs)
with definite and  increasing twists via   the scalar
product $p\mcdot x$ and the parameters $a_i$, $i=1,2,3$. The relations between
the two sets of DAs for the $N^\ast$, and for the scalar, pseudo-scalar, vector, axial-vector, and tensor DAs for nucleons are:
\bea
{\cal S}_1 \es S_1 \nnb \\
2 (p\mcdot x) \, {\cal S}_2 \es S_1-S_2 \nnb \\
 {\cal P}_1 \es P_1 \nnb \\
2 (p\mcdot x) \, {\cal P}_2 \es P_2-P_1 \nnb \\
\V_1 \es V_1 \nnb \\
2 (p\mcdot x) \, \V_2 \es V_1 - V_2 - V_3 \nnb \\
2 \V_3 \es V_3 \nnb \\
4 (p\mcdot x) \, \V_4  \es - 2 V_1 + V_3 + V_4  + 2 V_5 \nnb \\
4 (p\mcdot x) \V_5 \es V_4 - V_3 \nnb \\
4 (p\mcdot x)^2  \V_6 \es - V_1 + V_2 +  V_3 +  V_4 + V_5 - V_6 \nnb \\
\A_1 \es A_1 \nnb \\
2 (p\mcdot x) \A_2 \es - A_1 + A_2 -  A_3 \nnb \\
2 \A_3 \es A_3 \nnb \\
4 (p\mcdot x) \A_4 \es - 2 A_1 - A_3 - A_4  + 2 A_5 \nnb \\
4 (p\mcdot x) \A_5 \es A_3 - A_4 \nnb \\
4 (p\mcdot x)^2  \A_6 \es  A_1 - A_2 +  A_3 +  A_4 - A_5 + A_6 \nnb \\
\T_1 \es T_1 \nnb \\
2 (p\mcdot x) \T_2 \es T_1 + T_2 - 2 T_3 \nnb \\
2 \T_3 \es T_7 \nnb \\
 2 (p\mcdot x) \T_4 \es T_1 - T_2 - 2  T_7 \nnb \\
2 (p\mcdot x) \T_5 \es - T_1 + T_5 + 2  T_8 \nnb \\
4 (p\mcdot x)^2 \T_6 \es 2 T_2 - 2 T_3 - 2 T_4 + 2 T_5 + 2 T_7 + 2 T_8 \nnb \\
4 (p\mcdot x) \T_7 \es T_7 - T_8 \nnb \\
4 (p\mcdot x)^2 \T_8 \es -T_1 + T_2 + T_5 - T_6 + 2 T_7 + 2 T_8~ \nnb
\eea
where the terms in $x^2$,  $\mathcal{V}_1^M,\mathcal{A}_1^M$ and $\mathcal{T}_1^M$
are left aside.

The distribution amplitudes $F[a_i(p\mcdot x)]$=  $S_i$, 
$P_i$, $V_i$, $A_i$, $T_i$ can be represented as:
\bea\label{dependent1}
F[a_i (p\mcdot x)]=\int dx_1dx_2dx_3\delta(x_1+x_2+x_3-1) e^{ip\cdot
x\Sigma_ix_ia_i}F(x_i)~.\nnb
\eea
where, $x_{i}$ with $i=1,~2$ and $3$ are longitudinal momentum
fractions carried by the participating quarks.

The explicit expressions for the $\Lambda$ DAs up to twist 6 are given as:\\
Twist--$3$ DAs:
\bea 
\label{twist3}
V_1(x_i,\mu) \es 120 x_1 x_2 x_3 \left[\phi_3^0(\mu) + 
\phi_3^+(\mu) (1- 3 x_3)\right]\,,
\nnb \\
A_1(x_i,\mu) \es 120 x_1 x_2 x_3 (x_2 - x_1) \phi_3^-(\mu) ~,
\nnb \\
T_1(x_i,\mu) \es 120 x_1 x_2 x_3 
\Big[\phi_3^0(\mu) - \frac12\left(\phi_3^+ - \phi_3^-\right)(\mu) 
(1- 3 x_3)\Big]~. \nnb
\eea
Twist--$4$ DAs:
\bea
\label{twist4}
V_2(x_i,\mu)  \es 24 x_1 x_2 
\left[\phi_4^0(\mu)  + \phi_4^+(\mu)  (1- 5 x_3)\right] ~,\nnb\\
A_2(x_i,\mu)  \es 24 x_1 x_2 (x_2 - x_1) \phi_4^-(\mu) ~,\nnb \\
T_2(x_i,\mu) \es 24 x_1 x_2 \left[
\xi_4^0(\mu) + \xi_4^+(\mu)( 1- 5 x_3)\right]~,\nnb \\
V_3(x_i,\mu)  \es  12 x_3 \left[\psi_4^0(\mu)(1-x_3)
+ \psi_4^+(\mu)( 1-x_3 - 10 x_1 x_2)
+ \psi_4^-(\mu) (x_1^2 + x_2^2 - x_3 (1-x_3) ) \right]~,\nnb \\
A_3(x_i,\mu) \es 12 x_3 (x_2-x_1) 
\left[\left(\psi_4^0 + 
\psi_4^+ \right)(\mu)+ \psi_4^-(\mu) (1-2 x_3) \right] ~,\nnb \\
T_3(x_i,\mu)  \es 6 x_3 \left[
(\phi_4^0 + \psi_4^0 + \xi_4^0)(\mu)(1-x_3) +
(\phi_4^+ + \psi_4^+ + \xi_4^+)(\mu) ( 1-x_3 - 10 x_1 x_2)\right. \nnb \\
\ar \left.(\phi_4^- - \psi_4^- + \xi_4^-)(\mu) (x_1^2 + x_2^2 - x_3 (1-x_3) )\right],\nnb \\
T_7(x_i,\mu)  \es 6 x_3 \left[
(\phi_4^0 + \psi_4^0 - \xi_4^0)(\mu)(1-x_3) +
(\phi_4^+ + \psi_4^+ - \xi_4^+)(\mu) ( 1-x_3 - 10 x_1 x_2)\right. \nnb \\
\ar \left.(\phi_4^- - \psi_4^- - \xi_4^-)(\mu) (x_1^2 + x_2^2 - x_3 (1-x_3) )
\right]~,\nnb \\
S_1(x_i,\mu) \es 6 x_3 (x_2-x_1) \left[
(\phi_4^0 + \psi_4^0 + \xi_4^0 + \phi_4^+ + \psi_4^+ 
+ \xi_4^+)(\mu)+ (\phi_4^- - \psi_4^- + \xi_4^-)(\mu)(1-2 x_3) \right]~,\nnb \\
P_1(x_i,\mu) \es 6 x_3 (x_1-x_2) \left[
(\phi_4^0 + \psi_4^0 -\xi_4^0 + \phi_4^+ + \psi_4^+ - 
\xi_4^+)(\mu) + (\phi_4^- - \psi_4^- - \xi_4^-)(\mu)(1-2 x_3) \right]~.\nnb
\eea
Twist--$5$ DAs:
\bea
\label{twist5} 
V_4(x_i,\mu) \es 3 \left[\psi_5^0(\mu)(1-x_3) 
+ \psi_5^+(\mu)(1-x_3 - 2 (x_1^2 +  x_2^2))
+ \psi_5^-(\mu)\left(2 x_1x_2 - x_3(1-x_3)\right) \right]~,\nnb\\
A_4(x_i,\mu) \es 3 (x_2 -x_1)
\left[- \psi_5^0(\mu) + \psi_5^+(\mu)(1- 2 x_3) + \psi_5^-(\mu) x_3\right]~,\nnb \\
T_4(x_i,\mu) \es \frac32 \left[
(\phi_5^0 +  \psi_5^0 + \xi_5^0) (\mu)(1-x_3)
+ \left(\phi_5^+ + \psi_5^+ + \xi_5^+ \right)(\mu)(1-x_3 - 2 (x_1^2 +  x_2^2))
\right. \nnb \\
\ar \left. \left(\phi_5^- - \psi_5^- + \xi_5^- \right) (\mu)\left(2 x_1x_2 -
x_3(1-x_3)\right)\right]~,\nnb \\
T_8(x_i,\mu) \es \frac32 \left[
(\phi_5^0 +  \psi_5^0 - \xi_5^0) (\mu)(1-x_3)
+ \left(\phi_5^+ + \psi_5^+ - \xi_5^+ \right)(\mu)(1-x_3 - 2 (x_1^2 +  x_2^2))
\right. \nnb \\
\ar \left. \left(\phi_5^- - \psi_5^- - \xi_5^- \right) (\mu)\left(2 x_1x_2 - x_3(1-x_3)\right) \right]~, \nnb \\
V_5(x_i,\mu) \es 6 x_3 
\left[\phi_5^0(\mu)  + \phi_5^+(\mu)(1- 2 x_3)\right]~, \nnb\\
A_5(x_i,\mu) \es 6 x_3 (x_2-x_1) \phi_5^-(\mu) ~, \nnb\\
T_5(x_i,\mu) \es 6 x_3 \left[ \xi_5^0(\mu) + \xi_5^+(\mu)( 1- 2 x_3)\right]~,\nnb \\
S_2(x_i,\mu) \es \frac32 (x_2 -x_1) 
\left[- \left(\phi_5^0 + \psi_5^0 + \xi_5^0\right)(\mu)
+ \left(\phi_5^+ + \psi_5^+ + \xi_5^+\right)(\mu) (1- 2 x_3) \right. \nnb \\
\ar \left. \left(\phi_5^- - \psi_5^- + \xi_5^- \right)(\mu) x_3 \right]~, \nnb \\
P_2(x_i,\mu) \es \frac32 (x_2 -x_1)
\left[- \left(-\phi_5^0 - \psi_5^0 + \xi_5^0\right)(\mu)
+ \left(-\phi_5^+ - \psi_5^+ + \xi_5^+\right)(\mu) (1- 2 x_3)\right. \nnb \\
\ar \left. \left(-\phi_5^- + \psi_5^- + \xi_5^- \right)(\mu) x_3  \right]~. \nnb
\eea
Twist-6:
\bea
\label{twist6}
V_6(x_i,\mu) \es 2 \left[\phi_6^0(\mu) +  \phi_6^+(\mu) (1- 3 x_3)\right]~,
\nnb \\
A_6(x_i,\mu) \es 2 (x_2 - x_1) \phi_6^- ~, \nnb \\
T_6(x_i,\mu) \es 2 \Big[\phi_6^0(\mu) - 
\frac12\left(\phi_6^+-\phi_6^-\right) (1\!-\! 3 x_3)\Big]~.\nnb
\eea
Finally the $x^2$ corrections to the corresponding expressions 
${\cal V}_1^M$, ${\cal A}_1^M$, ${\cal T}_1^M$
for the leading twist DAs $V_1$, $A_1$ and $T_1$
in the momentum fraction space are given as:  
\bea
{\cal V}_1^M (x_2) \es \int \limits_0^{1-x_2} dx_1 
V_1^{M}(x_1, x_2,1-x_1-x_2) \nnb \\
\es \frac{x_2^2}{24} \left[ f_{N^\ast} C_{f}^u (x_2) + 
\lambda_1^{N^\ast} C_{\lambda}^u (x_2)\right]~,\nnb
\eea
where
\bea
C_{f}^u (x_2)\es (1 - x_2)^3 \Big[113 + 495x_2 - 552x_2^2 
- 10A_1^u(1 - 3x_2) \nnb \\
\ar  2V_1^d(113 - 951x_2 + 828x_2^2) \Big]~,\nnb \\
C_{\lambda}^u (x_2) \es - (1- x_2)^3 
              \Big[13 - 20f_1^d + 3x_2 + 10f_1^u(1 \!-\! 3x_2)\Big]~. \nnb
\eea
The expression for the axial--vector function ${\cal A}_1^{M(u)} (x_2)$ is given
as:
\bea
{\cal A}_1^{M(u)} (x_2) \es \int_0^{1-x_2} dx_1 A_1^{M}(x_1, x_2,1-x_1-x_2)~,\nnb\\ 
      \es \frac{ x_2^2}{24} (1 - x_2)^3 \left[ f_{N^\ast} D_{f}^u (x_2)  + 
\lambda_1^{N^\ast} D_{\lambda}^u(x_2) \right]~,\nnb 
\eea
with
\bea
D_{f}^u(x_2) \es 11 + 45 x_2 - 2 A_1^u (113 - 951 x_2 + 828 x_2^2 )
+ 10 V_1^d (1 - 30 x_2)~, \nnb \\
D_{\lambda}^u (x_2)& = &  29 - 45 x_2 - 10 f_1^u (7 - 9 x_2) - 
20 f_1^d (5 - 6 x_2)~. \nnb
\eea
Similarly, we get for the function ${\cal T}_1^{M(u)} (x_2)$:
\bea
{\cal T}_1^{M(u)} (x_2) \es \int_0^{1-x_2} dx_1 T_1^{M}(x_1, x_2,1-x_1-x_2)~,\nnb\\ 
\es \frac{ x^2}{48}  \left[ f_{N^\ast} E_{f}^u(x_2)  +
\lambda_1^{N^\ast} E_{\lambda}^u(x_2)  \right]~,\nnb
\eea
where
\begin{eqnarray}
E_{f}^u (x_2) \es -\Big\{
(1 - x_2) \Big[3 (439 + 71 x_2 - 621 x_2^2 + 587 x_2^3 
- 184 x_2^4) \nnb \\
\ar 4 A_1^u (1 - x_2)^2 (59 - 483 x_2 + 414 x_2^2) \nnb \\
\ek 4 V_1^d (1301 - 619 x_2 - 769 x_2^2 + 1161 x_2^3 - 414 x_2^4)\Big]
\Big\}
- 12 (73 - 220V_1^d) \ln[x_2]~, \nnb \\
E_{\lambda}^u (x_2) \es -\Big\{(1 - x_2) 
\Big[5 - 211 x_2 + 281 x_2^2 - 111 x_2^3 
+ 10 (1 + 61 x_2 - 83 x_2^2 + 33 x_2^3) f_1^d \nnb \\
\ek 40(1 - x_2)^2 (2 - 3 x_2) f_1^u\Big]
\Big\} - 12 (3 - 10 f_1^d) \ln[x_2]~.\nnb
\eea

The following functions are encountered to the above amplitudes and they
can be  defined in
terms of the  8 independent parameters, namely  $f_{N^\ast}$,
$\lambda_1$, $\lambda_2$ and $f_1^u,f_1^d,f_2^d,A_1^u,V_1^d$:\\

\bea
\label{DAs}
\phi_3^0 \es \phi_6^0 = f_{N^\ast} \nnb \\
\phi_4^0 \es \phi_5^0 = {1 \over 2} (f_{N^\ast} + \lambda_1^{N^\ast}) \nnb \\
\xi_4^0 \es \xi_5^0 = {1 \over 6} \lambda_2^{N^\ast} \nnb \\
\psi_4^0 \es \psi_5^0 = {1 \over 2} (f_{N^\ast} - \lambda_1^{N^\ast})~, \nnb\\
\phi_3^- \es {21 \over 2} f_{N^\ast} A_1^u,~~\phi_3^+ = {7 \over 2}
f_{N^\ast} (1 - V_1^d)~, \nnb \\
\phi_4^+ \es {1\over 4} \left[f_{N^\ast} (3-10 V_1^d) + \lambda_1^{N^\ast}
(3 - 10 f_1^d) \right]~, \nnb \\
\phi_4^- \es - {5\over 4} \left[f_{N^\ast} (1-2 A_1^u) - \lambda_1^{N^\ast}
(1 - 2 f_1^d - 4 f_1^u) \right]~, \nnb \\
\psi_4^+ \es - {1\over 4} \left[f_{N^\ast} (2 + 5 A_1^u - 5 V_1^d) - 
\lambda_1^{N^\ast} (2 - 5 f_1^d - 5 f_1^u) \right]~, \nnb \\
\psi_4^- \es {5\over 4} \left[f_{N^\ast} (2 - A_1^u - 3 V_1^d) - 
\lambda_1^{N^\ast} (2 - 7 f_1^d + f_1^u) \right]~, \nnb \\
\xi_4^+ \es {1\over 16} \lambda_2^{N^\ast} (4 - 15 f_2^d)~, \nnb \\
\xi_4^- \es {5\over 16} \lambda_2^{N^\ast} (4 - 15 f_2^d)~, \nnb \\
\phi_5^+ \es  - {5\over 6} \left[f_{N^\ast} (3 + 4 V_1^d) - \lambda_1^{N^\ast}
(1 - 4 f_1^d)\right]~, \nnb \\
\phi_5^- \es - {5\over 3} \left[f_{N^\ast} (1 - 2 A_1^u) -
\lambda_1^{N^\ast} (f_1^d - f_1^u)\right]~, \nnb \\
\psi_5^+ \es - {5\over 6} \left[f_{N^\ast} (5 + 2 A_1^u - 2 V_1^d) - 
\lambda_1^{N^\ast} (1 - 2 f_1^d - 2 f_1^u) \right]~,\nnb \\
\psi_5^- \es {5\over 3} \left[f_{N^\ast} (2 - A_1^u - 3 V_1^d) +     
\lambda_1^{N^\ast} (f_1^d - f_1^u) \right]~, \nnb \\
\xi_5^+ \es {5\over 36} \lambda_2^{N^\ast}  (2 - 9 f_2^d)~,\nnb \\
\xi_5^- \es -{5\over 4} \lambda_2^{N^\ast} f_2^d~, \nnb \\
\phi_6^+ \es  {1\over 2} \left[f_{N^\ast} (1 - 4 V_1^d) -
\lambda_1^{N^\ast} (1 - 2 f_1^d)\right]~, \nnb \\
\phi_6^- \es {1\over 2} \left[f_{N^\ast} (1 + 4 A_1^d) +                   
\lambda_1^{N^\ast} (1 - 4 f_1^d - 2 f_1^u)\right]~, \nnb
\eea
where the parameters $A_1^u,~V_1^d,~f_1^d,~f_1^u$, and $f_2^d$ are defined as
\cite{r21},
\bea
A_1^u \es \varphi_{10} + \varphi_{11}~,\nnb \\
V_1^d \es {1\over 3} - \varphi_{10} + {1\over 3} \varphi_{11}~,\nnb \\
f_1^u \es {1\over 10} - {1\over 6} {f_{N^\ast} \over \lambda_1^{N^\ast}} -
{3\over 5} \eta_{10} - {1\over 3} \eta_{11}~, \nnb \\
f_1^d \es {3\over 10} - {1\over 6} {f_{N^\ast} \over \lambda_1^{N^\ast}} +  
{1\over 5} \eta_{10} - {1\over 3} \eta_{11}~, \nnb \\
f_2^d \es {4\over 15} + {2\over 5} \xi_{10}~,\nnb
\eea
The numerical values of the parameters 
$\varphi_{10},~\varphi_{11},~\varphi_{20},~\varphi_{21},~\varphi_{22},
~\eta_{10},~\eta_{11}$, and $f_{N^\ast}/\lambda_1^{N^\ast}$,
and $\lambda_1^{N^\ast}/\lambda_1^N$ are presented in Table \ref{tab:data}
(this Table is taken from \cite{r21}).

\begin{table}[h]
\caption{Parameters of the DAs for the $N^\ast(1535)$ baryon at $\mu^2 =
	2~\mathrm{GeV}^2$}\label{tab:data}
\renewcommand{\arraystretch}{1.3}
\addtolength{\arraycolsep}{-0.5pt}
\small
$$
\begin{array}{|c|c|c|c|c|c|c|c|c|c|}
\hline \hline
\mbox{Model} & \lambda_1^{N^\ast}/\lambda_1^N &
f_{N^\ast}/\lambda_1^{N^\ast} & \varphi_{10} & \varphi_{11} & \varphi_{20} &
\varphi_{21} & \varphi_{22} & \eta_{10} & \eta_{11} \\  \hline
\mbox{ LCSR--1} & 0.633 & 0.027 & 0.36 & -0.95 & 0 & 0 & 0 & 0     & 0.94 \\
\mbox{LCSR--2}  & 0.633 & 0.027 & 0.37 & -0.96 & 0 & 0 & 0 & -0.29 & 0.23 \\
\hline \hline
\end{array}
$$

\renewcommand{\arraystretch}{1}
\addtolength{\arraycolsep}{-1.0pt}
\end{table}

\clearpage


\newpage

%
%
%
\section*{Figure captions}
\textbf{Fig. (1)} The dependence of the form-factor $ G _A(Q^2) $ on $ Q^2 $ for various values of $ \beta $ with the data set LCSR-1. \\ \\ 
\textbf{Fig. (2)} The same as Fig. (1), but with the LCSR-2.\\ \\
\textbf{Fig. (3)} The same as Fig. (1), but for $ G _A ^S (Q^2) $.\\ \\
\textbf{Fig. (4)} The same as Fig. (3), but with LCSR-2.\\ \\
\textbf{Fig. (5)} The same as Fig. (1), but for $ G _P (Q^2) $.\\ \\
\textbf{Fig. (6)} The same as Fig. (5), but with LCSR-2.\\ \\
\textbf{Fig. (7)} The same as Fig. (1), but for $ G _P ^S (Q^2) $.\\ \\
\textbf{Fig. (8)} The same as Fig. (7), but with LCSR-2.
\newpage
\begin{figure}
[H]
\centering
\includegraphics[height=.9\textheight]{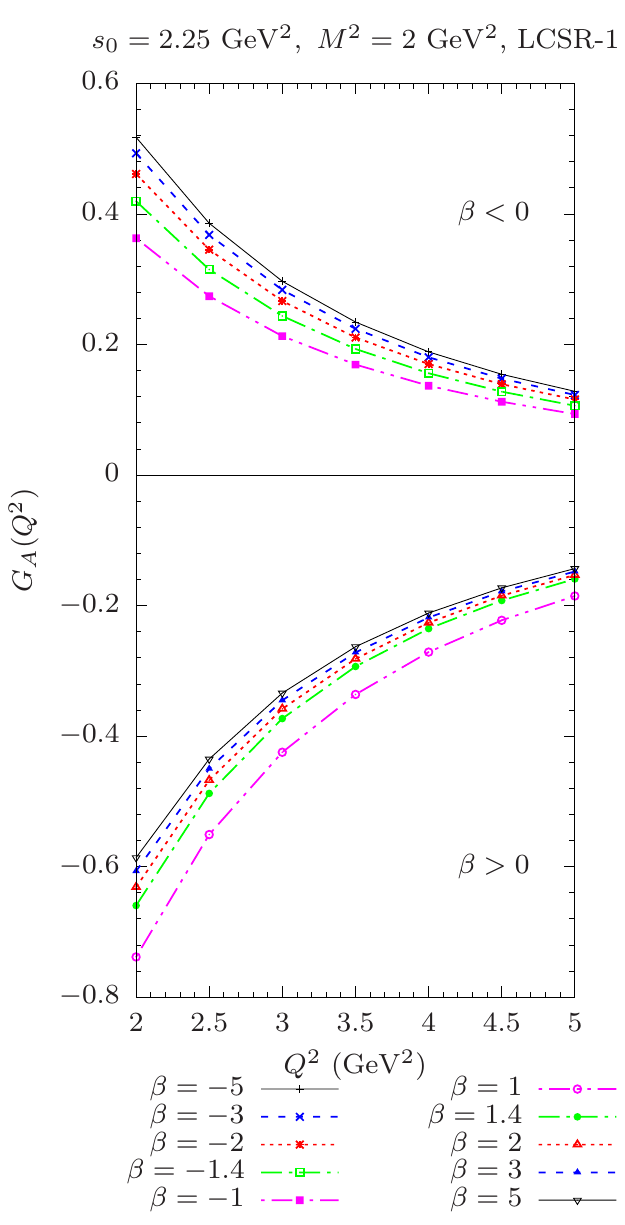}
\caption{}
\label{fig1}
\end{figure}

\begin{figure}
[H]
\centering
\includegraphics[height=.9\textheight]{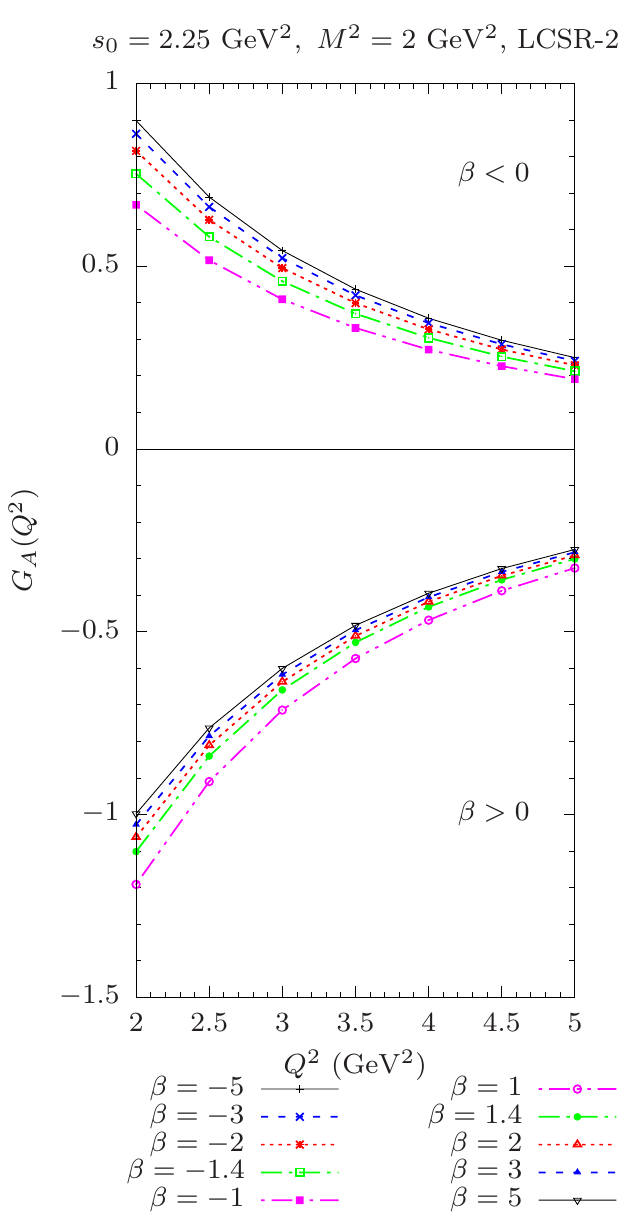}
\caption{}
\label{fig2}
\end{figure}

\begin{figure}
[H]
\centering
\includegraphics[height=.9\textheight]{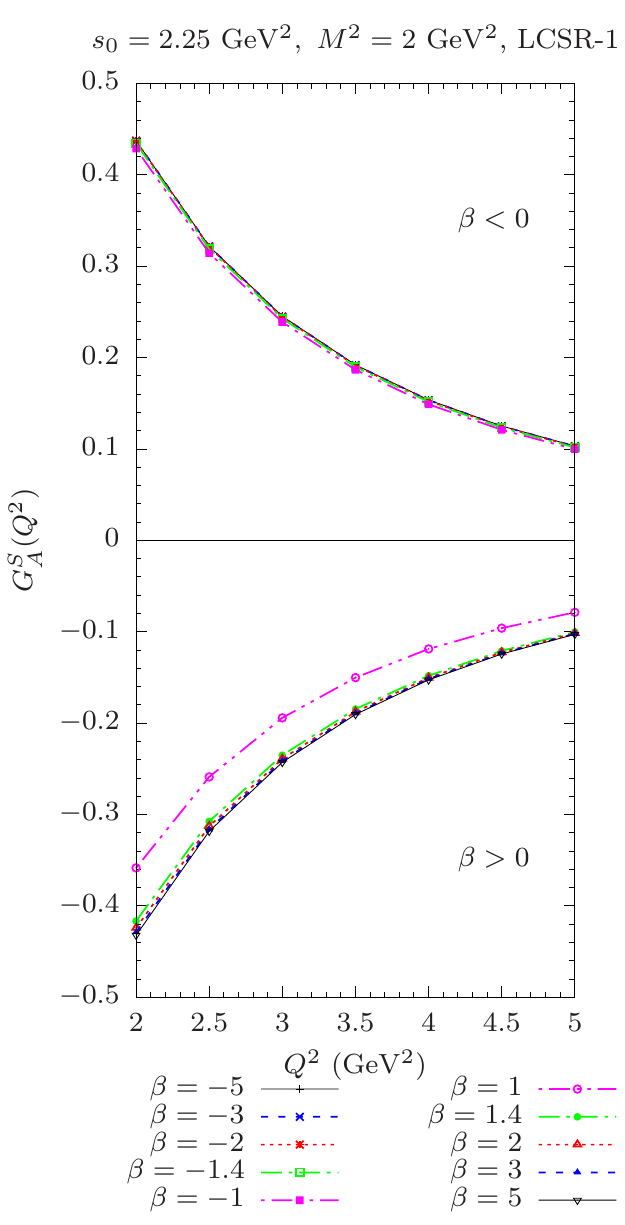}
\caption{}
\label{fig3}
\end{figure}

\begin{figure}
[H]
\centering
\includegraphics[height=.9\textheight]{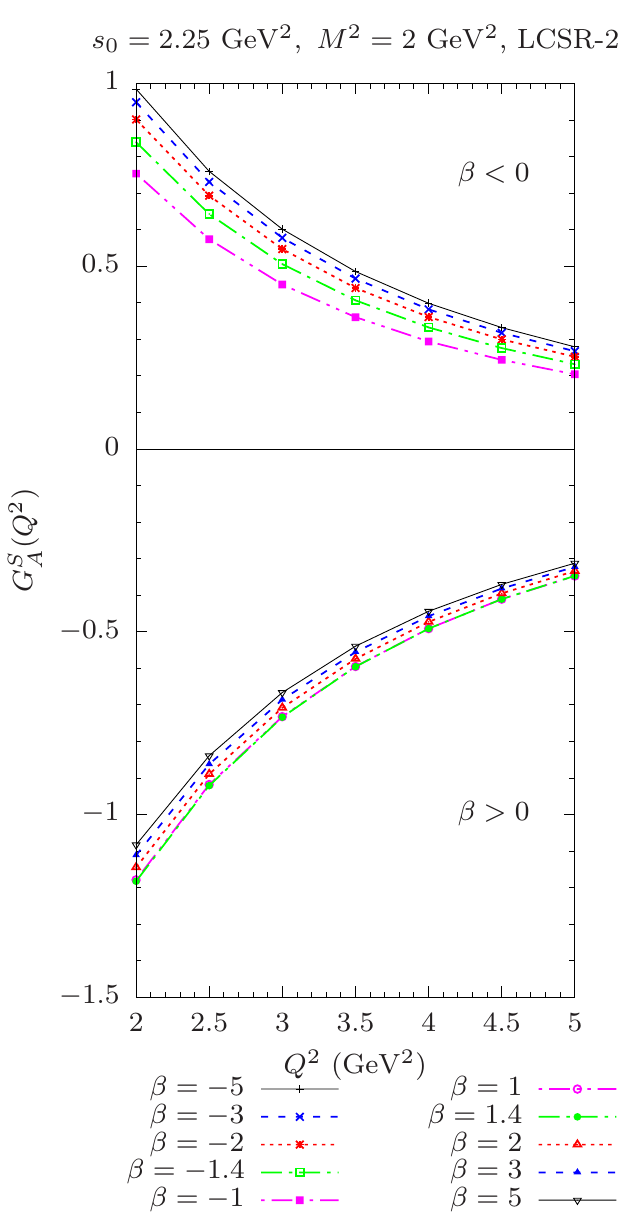}
\caption{}
\label{fig4}
\end{figure}

\begin{figure}
[H]
\centering
\includegraphics[height=.9\textheight]{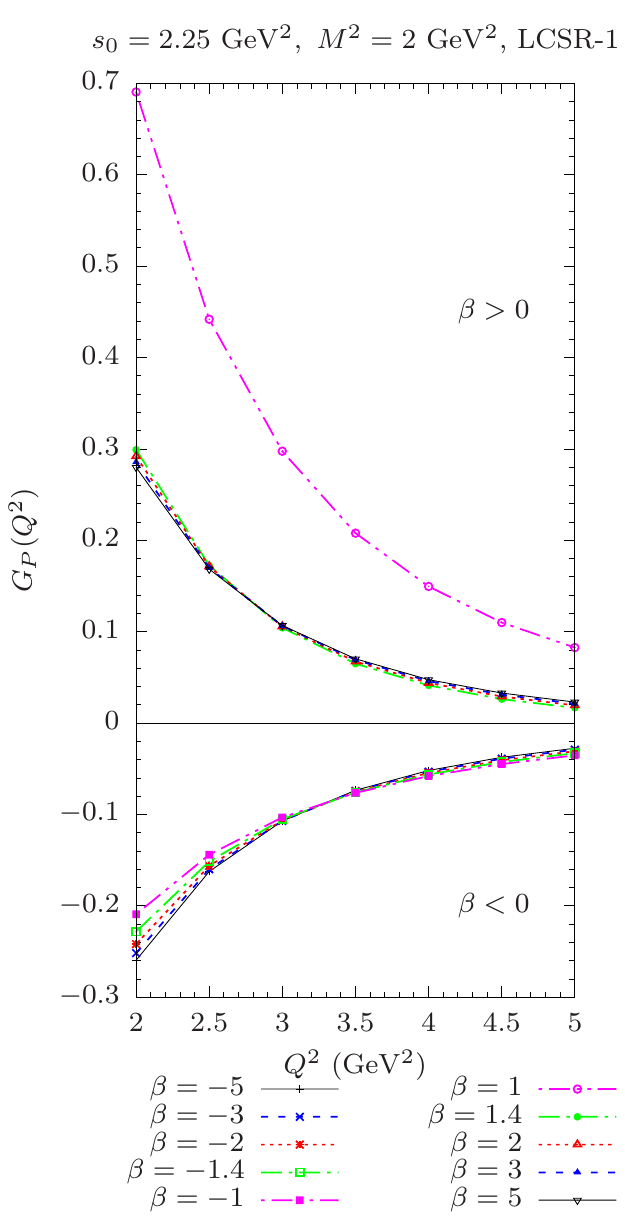}
\caption{}
\label{fig5}
\end{figure}

\begin{figure}
[H]
\centering
\includegraphics[height=.9\textheight]{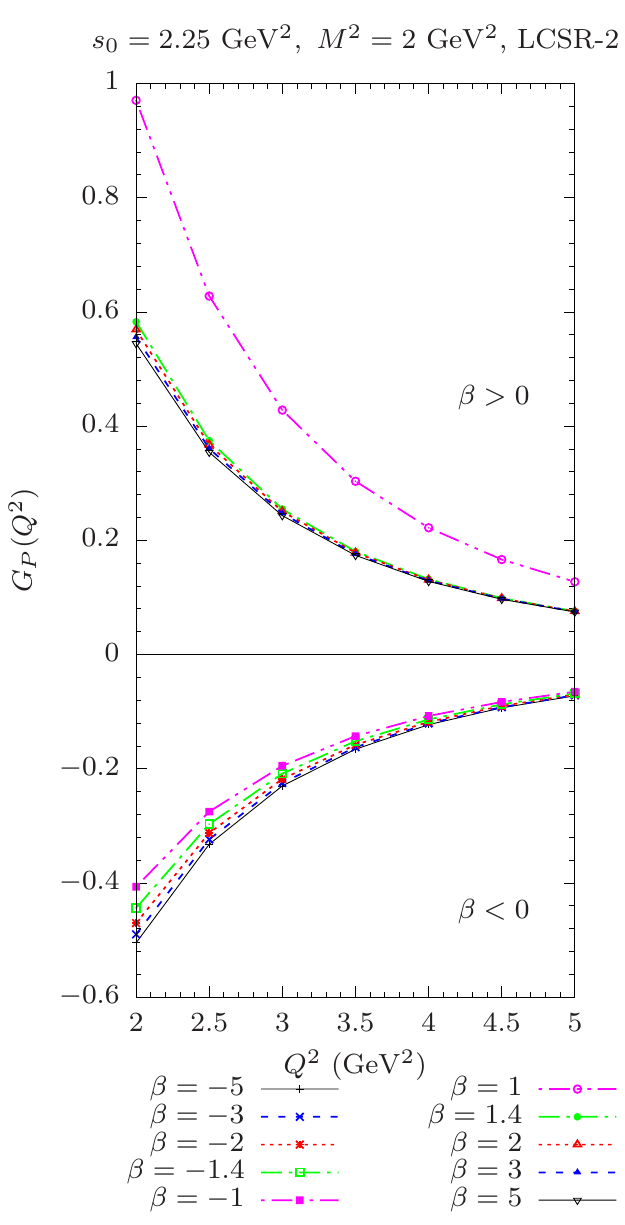}
\caption{}
\label{fig6}
\end{figure}

\begin{figure}
[H]
\centering
\includegraphics[height=.9\textheight]{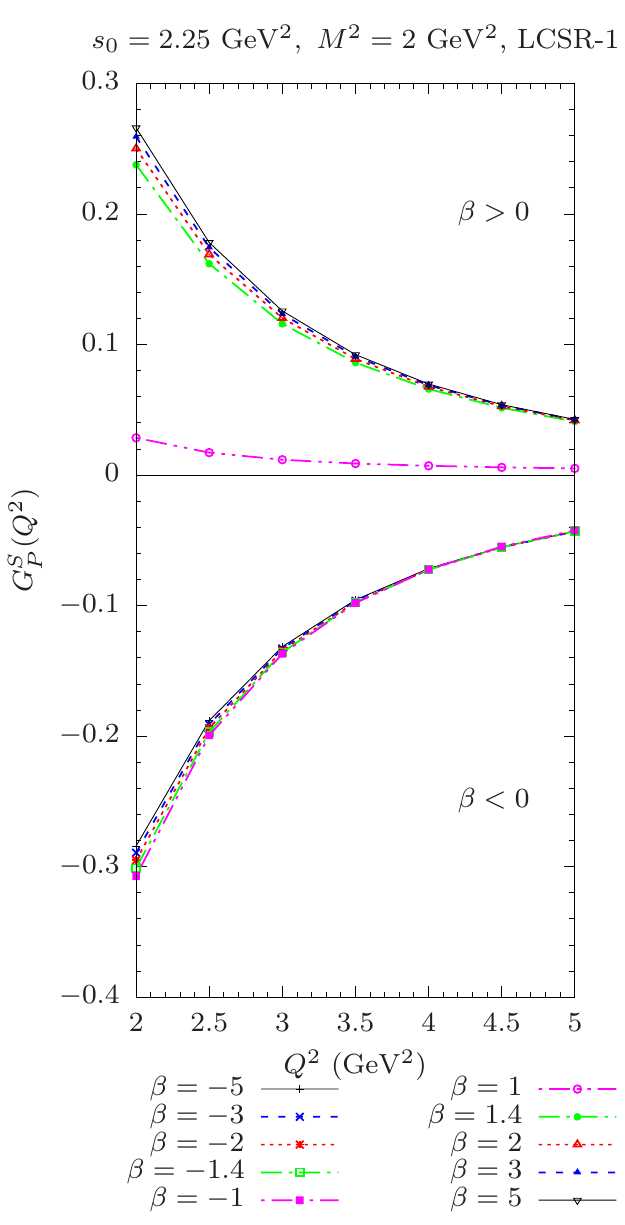}
\caption{}
\label{fig7}
\end{figure}

\begin{figure}
[H]
\centering
\includegraphics[height=.9\textheight]{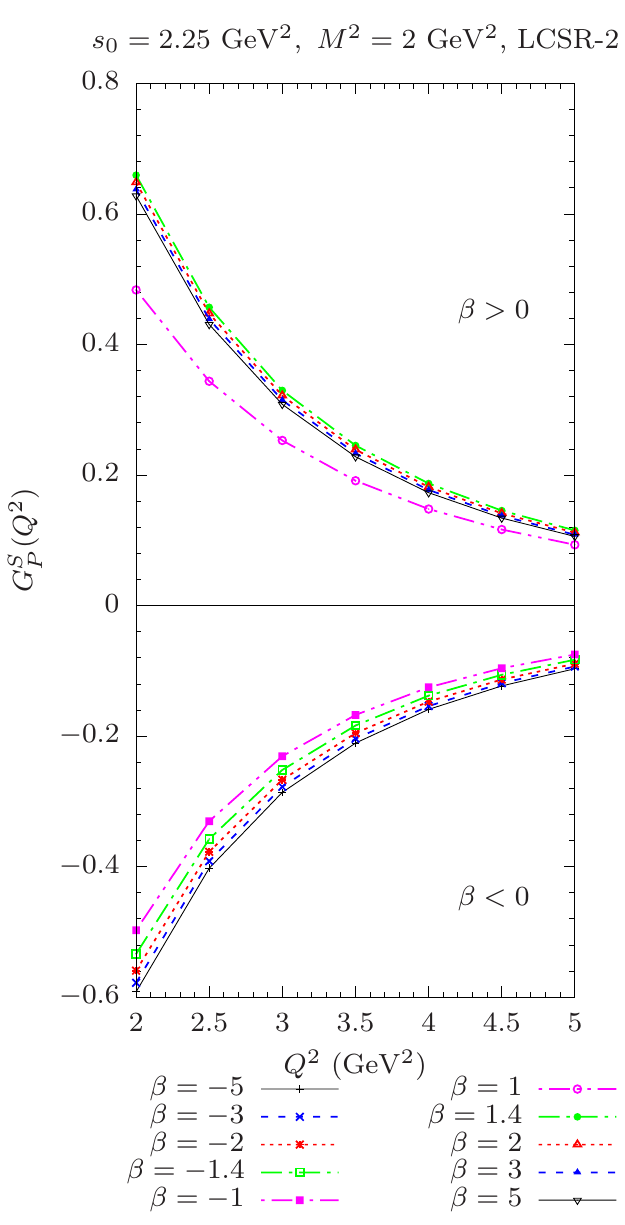}
\caption{}
\label{fig8}
\end{figure}

\end{document}